\documentclass[pre,reprint,amsmath,amssymb,superscriptaddress]{revtex4-2}
\usepackage{graphicx}
\usepackage{dcolumn}
\usepackage{bm}
\usepackage{amsmath}
\usepackage{epstopdf}
\usepackage{color}
\usepackage{verbatim}
\usepackage{caption}
\usepackage{subcaption}
\usepackage{float}

\begin{document}
\title{Mode-locking in a colloidal ring driven by power-modulated optical tweezers}
\author{Muyang Huang}
\affiliation{Department of Physics and Center for Complex Flow and Soft Matter Research, Southern University of Science and Technology, Shenzhen, Guangdong 518055, China}
\author{Pik-Yin Lai}
\email{pylai@phy.ncu.edu.tw}
\affiliation{Department of Physics and Center for Complex Systems, National Central University, Chung-Li District, Taoyuan City 320317, Taiwan}
\affiliation{Physics Division, National Center for Theoretical Sciences, Taipei 10617, Taiwan}
\author{Xiaoguang Ma}
\email{maxg@sustech.edu.cn}
\affiliation{Department of Physics and Center for Complex Flow and Soft Matter Research, Southern University of Science and Technology, Shenzhen, Guangdong 518055, China}
\begin{abstract}
Particles and clusters moving across real‑space periodic potentials can become locked to discrete directions or orientations due to competing symmetries. Here, we demonstrate an analogous locking phenomenon within a synthetic frequency space. We drive ring‑shaped colloidal clusters using a circular optical tweezer array, where power modulation of the traps generates coexisting, distinct potential waves. Relative displacements between the cluster and these waves trace zigzag trajectories across a synthetic two‑dimensional lattice, mirroring directionally locked motion in real‑space periodic potentials. By tuning the relative wave amplitudes, both the cluster’s direction in synthetic space and its velocity in real space exhibit discrete plateaus, both governed by square‑lattice symmetry. Furthermore, the formation of superlattices between the particles and potential wave minima mirrors the characteristic features of kinetically locked two‑dimensional clusters, demonstrating the capability to explore driven cluster dynamics within higher‑dimensional potentials using lower‑dimensional setups. Our findings establish new strategies for controlling transport of particle cluster via power‑modulated laser tweezers.
\end{abstract}
\maketitle
Kinetic locking and mode locking are two distinct classes of nonlinear transport phenomena in driven dynamical systems. Kinetic locking refers to the phenomenon where particles driven across two- and three-dimensional (2D and 3D) periodic potentials exhibit motion aligned not with the driving force, but with the potential landscape's discrete symmetry directions, which form a Devil's staircase \cite{1999-Reichhardt-Phase-PRLa,2001-Pierre-Louis-Oscillatory-PRL,2002-Korda-Kinetically-PRL,2003-MacDonald-Microfluidic-N,2004-Gopinathan-Statistically-PRLa,2004-Huang-Continuous-S,2004-Reichhardt-Directional-PRE,2007-Roichman-Colloidal-PRE,2009-Balvin-Directional-PRLa,2012-Bohlein-Experimental-PRLa,2015-Ma-SoftMatter,2015-Ma-PhysRevE.91.042306,2017-Ma-PhysRevE.96.012601,2022-Stuhlmuller-Colloidal-PRE}. This effect is ubiquitous, spanning adatoms on crystal surfaces \cite{2001-Pierre-Louis-Oscillatory-PRL}, electrons in charge density waves \cite{2001-Wiersig-Devils-PRL}, vortices in superconducting arrays \cite{1999-Reichhardt-Phase-PRLa, 2005-Togawa-Direct-PRL}, colloids in periodic potentials \cite{2002-Korda-Kinetically-PRL,2003-MacDonald-Microfluidic-N,2007-Roichman-Colloidal-PRE,2020-Stoop-Collective-PRL,2022-Stuhlmuller-Colloidal-PRE}, and grains sedimenting through periodic obstacles \cite{2004-Huang-Continuous-S,2009-Balvin-Directional-PRLa}. Recent work extends single-particle kinetic locking to 2D particle assemblies where collective effects give rise to novel transport behavior such as kinks/anti-kinks and orientational locking \cite{2012-Bohlein2012,2012-Bohlein-Experimental-PRLa,2019-Cao-Orientational-NP,2019-Tierno-Moire-NP,2020-Stoop-Collective-PRL}. Beyond fundamental interests, kinetic locking has enabled sorting and fragmentation technologies for microparticles and biomolecules \cite{2003-MacDonald-Microfluidic-N,2004-Huang-Continuous-S}, and holds the potential to guide particle cluster assembly for bottom-up manufacturing \cite{2019-Cao-Orientational-NP}.

Unlike kinetic locking, mode locking requires external periodic drives and arises when a system's internal frequency synchronizes with external ones. The defining feature of mode locking is quantized plateaus in velocity-force or voltage-current relationships, known as Shapiro steps after their first observation in microwave-driven Josephson junctions \cite{1963-PhysRevLett.11.80,1968-PhysRev.169.397}. Although numerous models have predicted mode-locked Shapiro steps for particles, vortices, and clusters \cite{2001-PhysRevLett.86.4112,2002-PhysRevLett.89.024101,2006-PhysRevLett.97.056101,2007-PhysRevLett.98.148001,2007-PhysRevLett.99.206101}, experimental realization remains challenging. To this end, the precise engineering of optical potentials using modern laser tweezers affords unparalleled control over microparticle transport, enabling to realize both integer and fractional Shapiro steps at the single-particle level \cite{2015-Juniper-Microscopic-NC,2025-Stikuts-Engineering-N}. Despite these advances, mode locking is still treated as a synchronization phenomenon, distinct from kinetic locking---an effect determined by lattice symmetry---leaving the fundamental relationship between the two an open question. 

Here, we investigate the driven dynamics of a one-dimensional (1D) ring-shaped colloidal cluster in a circular laser tweezers array. A unique feature of our system is the power modulation of individual tweezers, which generates a set of potential waves; the number and physical properties of these waves are engineered by tuning the modulation strength, modulation period, and the number of particles and tweezers. Driven by these waves, the colloidal ring exhibits rich dynamical behavior from single-wave trapping to inter-wave transitions, in excellent agreement with our theoretical and numerical predictions.

By tuning the relative amplitudes of two coexisting potential waves, the colloidal ring undergoes a non-smooth transition from trapping by one wave to the other: the ring's mean rotation speed vs modulation strength exhibits rational Shapiro steps. When characterized by its relative displacements to the two waves, the ring traces zigzag trajectories in a {\it synthetic} 2D lattice, where the mean directions of these trajectories form a Devil's staircase. These results establish an equivalence between real-space mode locking and synthetic-space directional locking, and provide a geometric interpretation of mode locking. Furthermore, during inter-wave transitions, 1D superlattices of potential minima emerge between the particles and distinct waves, mirroring the characteristics of kinetically locked 2D clusters \cite{2019-Cao-Orientational-NP}. This analogy unveils a pathway to studying higher-dimensional kinetic locking in clusters using lower-dimensional systems. By realizing both mode locking and kinetic locking in a single driven system, we demonstrate the versatility of power-modulated laser tweezers for probing and controlling nonequilibrium driven dynamics in soft condensed matter.
    
\section*{Results}
\subsection*{Creation of multiple potential waves}
\begin{figure}[h]
	\centering
	\includegraphics[width=0.95\columnwidth]{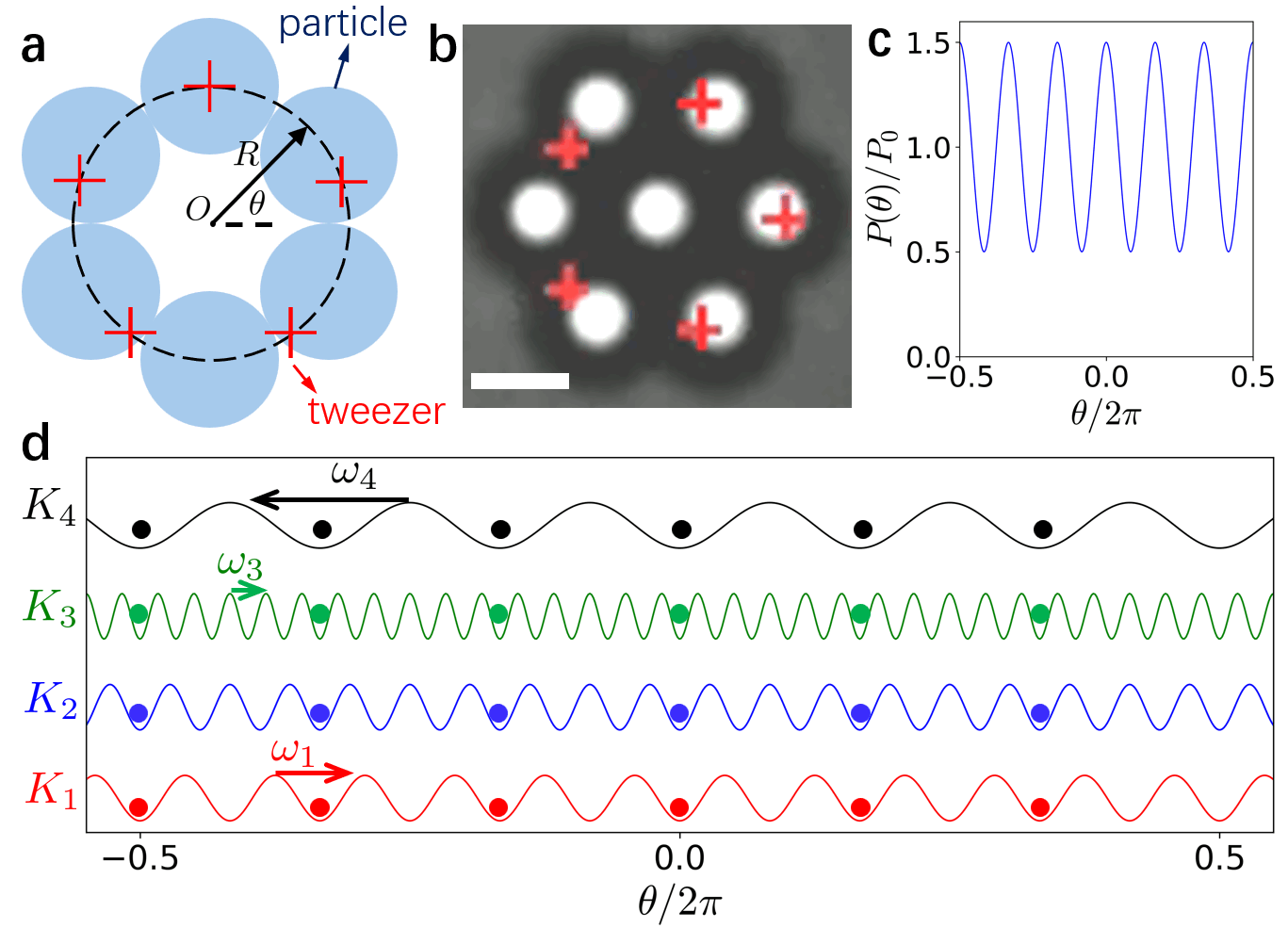}
	\caption{{\bf Colloidal ring and optical potentials.} {\bf a} A schematic of $N=6$ particles (blue circles) and $M=5$ laser tweezers (red crosses). $O$: center. $R$: radius. $\theta$: angular position. {\bf b} Sample snapshot. Bright blobs: particles. Red crosses: tweezers. Scale bar: 2 $\mu$m. {\bf c} Example of power modulation: $L=6$ and $\epsilon=0.5$. {\bf d} Waves in Eq.(\ref{eq:potential}) for $\{M,L\}=\{12,18\}$. Arrows denote $\omega_i$ ($i=1,3,4$) with lengths proportional to $|\omega_i|$. Dots denote superlattices of potential minima that overlap with the particles.}
	\label{fig:1}
\end{figure}

The ring-shaped colloidal cluster comprises $N=6$ polystyrene beads (nominal diameter $D\simeq 2.5$-$\mu$m) (Fig.~\ref{fig:1}a,b; the central particle does not contribute to the driven dynamics and is not counted). The cluster is confined in a quasi-2D aqueous environment and bound together by attractive depletion forces \cite{2016-Ma-PhysRevE.94.042606,2019-Ma-10.1063/1.5091564,2021-Ma-10.1063/5.0059084,2023-Ma-10.1063/5.0146155}. We drive the ring's rotation using a concyclic circular array of $M$ laser tweezers (red crosses in Fig.~\ref{fig:1}a,b) rotating at an angular speed $\omega$. We modulate the power of each tweezers as $P=P_0[1+\epsilon\cos(L\theta)]$, where $P_0\simeq 7$ mW is the mean power, $L$ is the modulation period, $0<\epsilon<1$ is the modulation strength, and $\theta$ denotes angular position (see Fig.~\ref{fig:1}c for an example). See the Methods Section and Supplementary Information (SI) for experimental details.

When $M\neq L$, the potential $u(\theta,t)$ experienced by a particle at time $t$ and position $\theta$ is approximated by \cite{2016-Egelhaaf-PhysRevA,2016-Juniper-PhysRevE,2021-Gieseler-Optical-AOP,supp} (see the Methods Section):
\begin{align}
&u(\theta, t) = -b_1\cos[K_1(\omega_1 t-\theta)]\nonumber -\epsilon b_2\cos[K_2(\omega_2 t-\theta)]\\
&  -\epsilon b_3\cos[K_3(\omega_3 t-\theta)]- \epsilon b_4\cos[K_4(\omega_4 t-\theta)].
\label{eq:potential}
\end{align}
$u(\theta,t)$ comprises four potential waves with wavenumbers $\{K_i\}$ and angular speeds $\{\omega_i\}$ ($i=1,2,3,4$) (Table~\ref{tab:amplitudes}). As an example, Fig.~\ref{fig:1}d schematically plots the waves for $\{M,L\}=\{12,18\}$ at $t=0$. In general, the $K_1$ wave travels at the same speed as the tweezers ($\omega_1=\omega$); $K_2$ is static ($\omega_2=0$); $K_3$ is slower than the tweezers [$\omega_3/\omega=M/(M+L)<1$]. The $K_4$ wave, however, can move faster in the same or opposite direction depending on the sign of $M-L$. For the particular example $\{M,L\}=\{12,18\}$, the $K_4=-6$ wave travels opposite to the tweezers. The factors $\epsilon b_i$ for the $K_2$, $K_3$, and $K_4$ waves are proportional to the modulation strength $\epsilon$; when $\epsilon=0$ (no modulation), only $K_1$ exists \cite{2007-Roichman-Colloidal-PRE}. The wavenumber-dependent coefficients $b_i\propto \exp(-CK_i^2)$ (see the Methods Section) imply that larger-wavenumber waves have negligible amplitudes.  

\begin{table*}[ht]
\caption{\label{tab:amplitudes}
Wavenumbers $\{K_i\}$, angular speeds $\{\omega_i\}$, and amplitudes $\{a_i\}$ of the waves in Eqs.~(\ref{eq:potential})--(\ref{eq:eom}).}
\begin{ruledtabular}
	\begin{tabular}{cccc}
		$i$ & $K_i$ & \(\omega_i/\omega\) & \( a_i \)  \\ 
		\colrule
		1 & $M$  & \(1\) & \(b_1 NM \) \\
		2 & $L$  & \(0 \) &\( \epsilon b_2 NL \) \\
		3 & $M+L$  & \(M/(M+L)\) & \(\epsilon b_3N(M+L)\)\\
		4 & $M-L$  & \(M/(M-L)\) & \(\epsilon b_4N(M-L)\)\\
	\end{tabular}
\end{ruledtabular}
\end{table*}

\subsection*{Activation of zero, one, or multiple waves}
The total potential experienced by the ring cluster is the sum of the individual particle potentials. Within the ring, the phase offset $j2\pi(K_i/N)$ of the $j$-th particle ($j=1,2,\dots,N$) dictates that the $K_i$ wave persists in the summation only if $K_i$ is divisible by $N$ (i.e., $K_i/N\in\mathbb{Z}$). Thus, the total potential becomes:
\begin{align}
	U(\theta,t) = -N\{&\sigma(K_1)b_1\cos[K_1(\omega_1 t-\theta)]\nonumber \\
	+&\sigma(K_2)b_2\cos(-K_2\theta)\nonumber \\
	+&\sigma(K_3)b_3\cos[K_3(\omega_3 t-\theta)]\nonumber \\
	+&\sigma(K_4)b_4\cos[K_4(\omega_4 t-\theta)]\},
	\label{eq:total_potential}
\end{align}
where the $\sigma$ function enforces the {\it divisibility condition}: $\sigma(x)=1$ if $x/N\in \mathbb{Z}$ and $\sigma(x)=0$ otherwise. For the example $\{M,L\}=\{12,18\}$, all four waves coexist (Fig.~\ref{fig:1}d). To illustrate this divisibility condition, we mark the wave minima that can trap the $N=6$ particles by colored dots in Fig.~\ref{fig:1}d; each group of same-color dots marks a superlattice of potential traps within the $K_i$ wave, with the superlattice spacing covering $K_i/N$ potential barriers. 

For general $\{M,L\}$, however, only a subset (or none) of the waves satisfy the divisibility condition and drive the ring. Specifically, the $\{\sigma(K_i)\}$ set defines seven distinct cases, five of which involve zero or only one wave in Eq.~(\ref{eq:total_potential}): $\{\sigma(K_i)\}=\{0,0,0,0\}$ (no waves), $\{1,0,0,0\}$ (only $K_1$), $\{0,1,0,0\}$ (only $K_2$), $\{0,0,1,0\}$ (only $K_3$), and $\{0,0,0,1\}$ (only $K_4$). There are two cases where multiple waves coexist: for $\{\sigma(K_i)\}=\{0,0,1,1\}$, the smaller-wavenumber $K_4$ wave always dominates over the $K_3$ wave; for $\{1,1,1,1\}$, wave amplitudes are tunable via $\{M,L\}$ and $\epsilon$. The divisibility criterion thus enables a novel suite of strategies for controlling the cluster's dynamics via the number of driving waves and wave properties tuned by $\{N,M,L\}$ and $\epsilon$, which distinguishes our technique from prior optical tweezers-based approaches \cite{2015-Juniper-Microscopic-NC,2025-Stikuts-Engineering-N}. 

\subsection*{Single-wave trapping}
\begin{figure}[h]
	\centering
	\includegraphics[width=0.95\columnwidth]{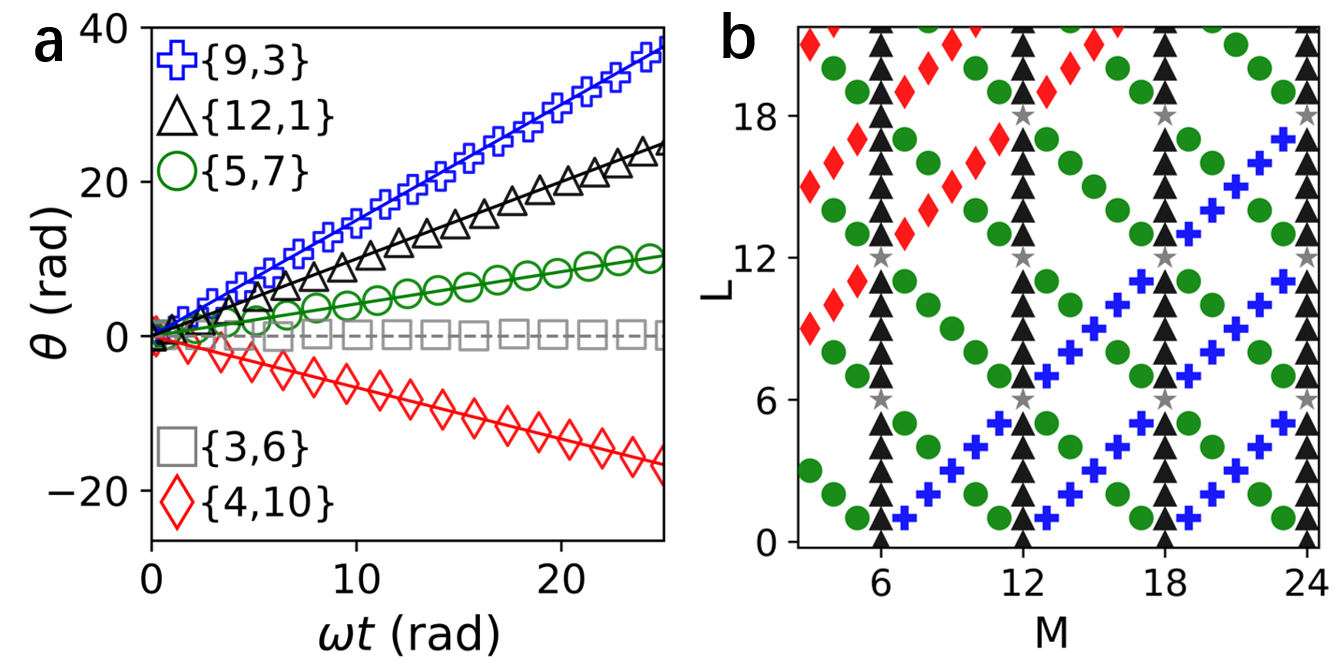}
	\caption{{\bf Single-wave trapping results.} {\bf a} Experimental results of $\theta(t)$ for $\epsilon=1/3$ and various $\{M,L\}$ cases, showing trapping by different waves: $K_1$ (black triangles), $K_2$ (grey squares), $K_3$ (green circles), and $K_4$ for $M>L$ (blue cross) and $M<L$ (red diamonds). {\bf b} Simulations of single-wave trapping by $K_1$ (black triangles), $K_3$ (green circles), and $K_4$ (blue cross for $M>L$ and red diamonds for $M<L$). Grey stars denote multi-wave cases.}
	\label{fig:singleWave}
\end{figure}

We use a slow speed $\omega=0.05$ rad/s for the tweezers to demonstrate single-wave trapping under weak viscous drag. Figure~\ref{fig:singleWave}a shows experimental results of the ring's angular displacement $\theta(t)$ for selected $\{M,L\}$ sets such that no wave or only one wave exists. The modulation strength is $\epsilon=1/3$. The average ring speed is computed by $\omega_p\equiv \lim_{t\to\infty}[\theta(t)-\theta(0)]/t$. Among these examples, $\{M,L\}=\{9,3\}$ (only $K_4$ exists) yields $\omega_p/\omega\simeq 3/2$, showing the ring rotates faster than the tweezers (Supplementary Movie 1). In contrast, $\{4,10\}$ yields $\omega_p/\omega\simeq -2/3$ (only $K_4$ exists), meaning opposite rotation to the tweezers (Supplementary Movie 2). We also observe ring rotation slower than the tweezers ($\omega_p/\omega\simeq 5/12$ for $\{M,L\}=\{5,7\}$ where only $K_3$ exists), equal speed ($\omega_p=\omega$ for $\{12,1\}$ where only $K_1$ exists), and no rotation ($\omega_p\simeq 0$ for $\{3,6\}$ where only $K_2$ exists) (Supplementary Movies 3--5), all consistent with the predictions (Table~\ref{tab:amplitudes}). We note that, despite the excellent agreement between the experiment and model, the concyclic configuration of the colloidal ring and tweezers array assumed in the model is only approximate: in certain cases (e.g., $\{M,L\}=\{5,7\}$), the centers of the two do not exactly overlap and exhibit a complex dynamical pattern (see Supplementary Movie 3), which highlights the inherently 2D nature of our experiment.

Figure~\ref{fig:singleWave}b compiles trapping waves observed in Brownian dynamics simulations (see the Methods Section), all in excellent agreement with experimental data and theoretical predictions. Note that for $\{M,L\}$ sets where multiple waves coexist (grey stars in Fig.~\ref{fig:singleWave}b), the ring's dynamics depend on $\epsilon$, which is discussed later. 

\begin{figure}[h]
	\centering
	\includegraphics[width=0.95\columnwidth]{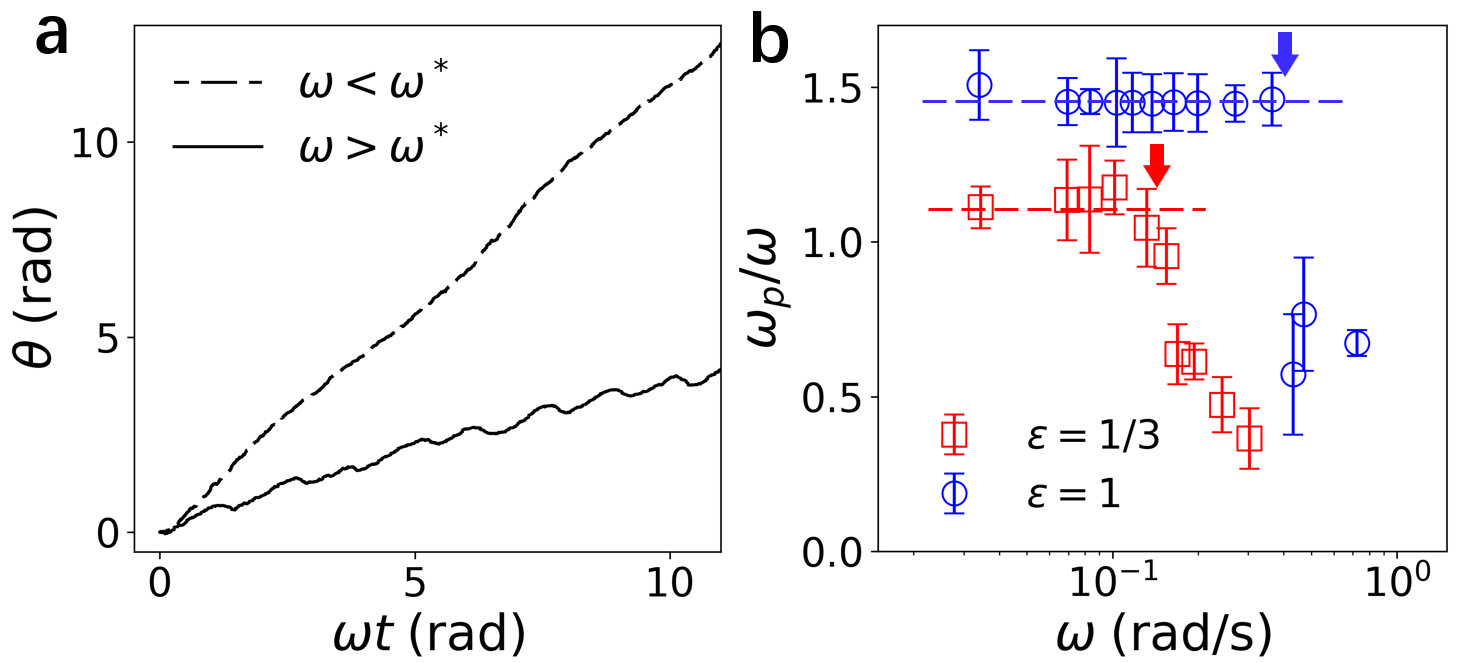}
	\caption{{\bf Critical trapping condition.} {\bf a} $\theta(t)$ below ($\omega=0.36$ rad/s, dashed line) and above ($\omega=0.72$ rad/s, solid line) $\omega^*\simeq 0.4$ rad/s for $\epsilon=1$. {\bf b} $\omega_p/\omega$ vs $\omega$ for $\epsilon=1/3$ (red squares) and $\epsilon=1$ (blue circles). The $\epsilon=1$ dataset is shifted upward by $0.3$ for clarity. Dashed lines denote $\omega_p/\omega=7/6$. Red (blue) arrow indicates the theoretical predictions of $\omega^*=0.13$ ($\omega^*=0.4$) rad/s for $\epsilon=1/3$ ($\epsilon=1$).}
	\label{fig:stability_exp}
\end{figure}

When the tweezers speed exceeds a critical value $\omega^*$ (see the Methods Section), the trapping wave can no longer synchronously transport the cluster against viscous drag. For demonstration, we use $\{M,L\}=\{7,1\}$, where only the $K_4=6$ wave persists. Figure~\ref{fig:stability_exp}a shows $\theta(t)$ for tweezers speeds below ($\omega=0.36$ rad/s) and above ($\omega=0.72$ rad/s) the critical value $\omega^*\simeq 0.40$ rad/s for $\epsilon=1$. When $\omega<\omega^*$, $\theta(t)$ increases smoothly with time, and $\omega_p/\omega=7/6$ matches the prediction. In contrast, $\omega>\omega^*$ leads to periodic ``stick-slip'' dynamics: ``stick'' events correspond to the ring climbing potential barriers, while ``slip'' events involve the ring falling to downstream potential minima within the wave (Supplementary Movie 6).

Figure~\ref{fig:stability_exp}b plots $\omega_p/\omega$ vs $\omega$ for two $\epsilon$ values. For $\epsilon=1$ (as in Fig.~\ref{fig:stability_exp}a), $\omega_p/\omega$ remains $7/6$ up to $\omega\simeq \omega^*$, indicating stable trapping. Above $\omega^*$, $\omega_p/\omega$ drops abruptly, indicating the ring is forced by viscous drag to hop between potential traps within the wave. For the smaller $\epsilon=1/3$, $\omega_p/\omega\simeq 7/6$ persists up to a smaller critical speed $\omega^*\simeq 0.13$ rad/s, confirming the linear dependence of $\omega^*$ on $\epsilon$ via $a_i$ (see Table~\ref{tab:amplitudes}).

\subsection*{Transitions between multiple waves}
We reuse the example $\{M,L\}=\{12,18\}$ to demonstrate the ring's dynamics when multiple waves coexist. Among the four waves, the amplitudes of $K_2=18$ and $K_3=30$ are much smaller than those of $K_1=12$ and $K_4=-6$ (due to their larger wavenumbers). Thus, the ring is effectively driven by $K_1$ and $K_4$. Notice that the amplitude of $K_4$ is tunable via $\epsilon$ (Table~\ref{tab:amplitudes}), such that $K_1$ is the dominant wave for $\epsilon\ll\epsilon^*\simeq 0.28$, while $K_4$ dominates for $\epsilon\gg\epsilon^*$; for $\epsilon\simeq \epsilon^*$, their amplitudes are comparable.

\begin{figure}[h]
	\centering
	\includegraphics[width=.95\columnwidth]{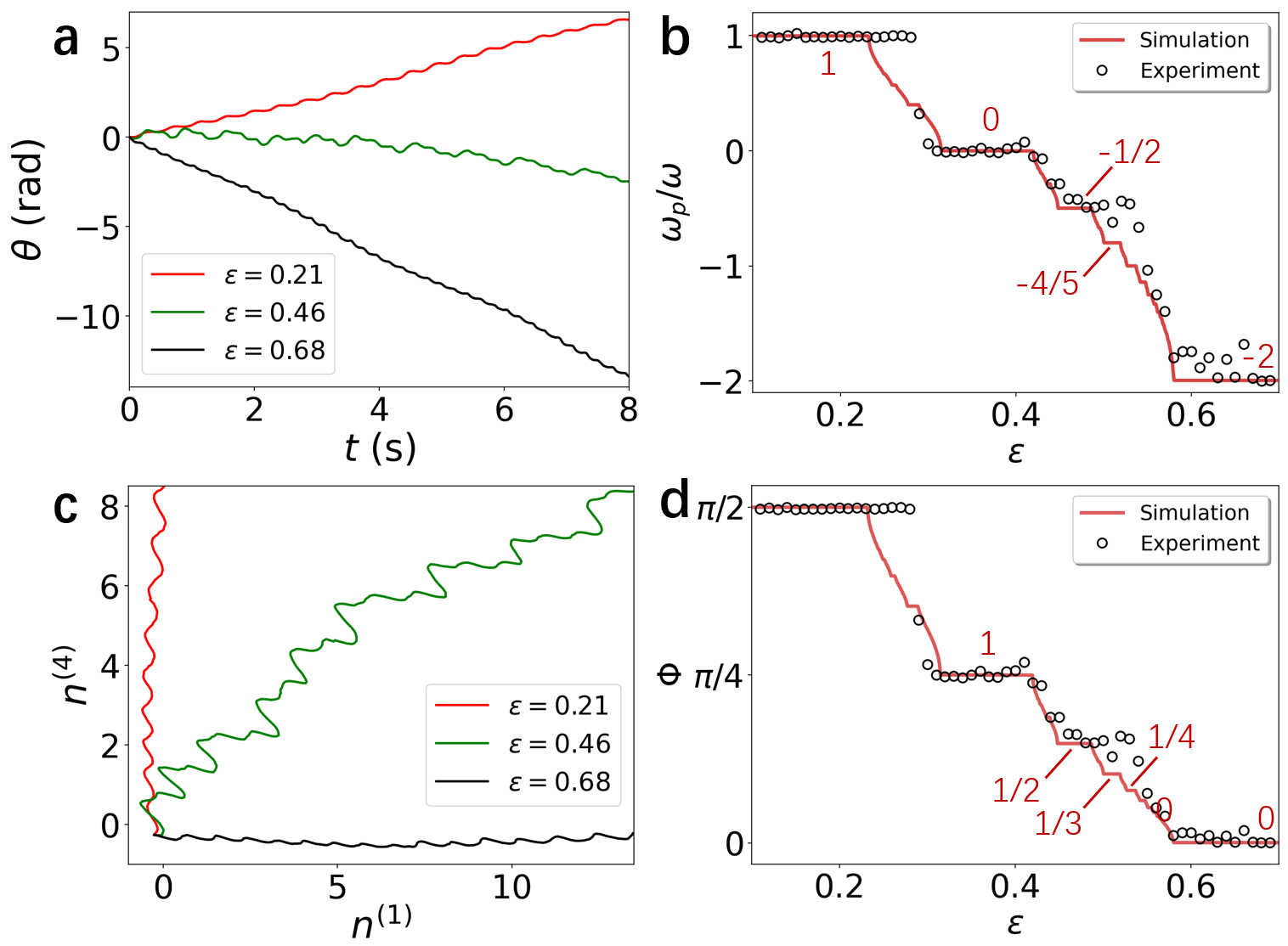}
	\caption{{\bf Mode locking in real space and directional locking in synthetic space.} {\bf a} $\theta(t)$ curves for $\epsilon=0.21$ (red), $0.46$ (green), and $0.68$ (black). {\bf b} $\omega_p/\omega$ vs $\epsilon$ for experiments (black circles) and simulations (red line). {\bf c} $n^{(4)}$ vs $n^{(1)}$ from the data in {\bf a}. {\bf d} $\Phi$ vs $\epsilon$ for experiments (black circles) and simulations (red line). Primary plateaus of $\omega_p/\omega$ and $\tan\Phi$ are labeled by corresponding rational numbers in {\bf b} and {\bf d}, respectively.}
	\label{fig:wave_transition}
\end{figure}

We select an intermediate tweezers speed $\omega=0.84$ rad/s to demonstrate transitions between states trapped by $K_1$ and $K_4$. Figure~\ref{fig:wave_transition}a shows $\theta(t)$ for three $\epsilon$ values: for $\epsilon=0.21$ ($0.68$), the ring is trapped by $K_1$ ($K_4$) (Supplementary Movies 7,8). The two $\theta(t)$ curves exhibit small-amplitude, rugged fluctuations caused by perturbative effects from the non-dominant wave. At the intermediate $\epsilon=0.46$, the ring's dynamics alternate between being trapped by $K_1$ and $K_4$, suggesting the two waves are competing for dominion (Supplementary Movie 9).

\subsection*{Mode locking in real space}
Figure~\ref{fig:wave_transition}b plots $\omega_p/\omega$ vs $\epsilon$, revealing a transition from $K_1$- to $K_4$-dominance induced by increasing the modulation strength. This transition is not smooth, exhibiting rational plateaus ($1,\dots,0,\dots,-2$) in ring speeds. Unlike previous studies where external periodic driving and internal temporal modulation are pre-defined \cite{2015-Juniper-Microscopic-NC,2025-Stikuts-Engineering-N}, the two frequencies in our experiment are both owing to external driving and their roles are switchable by $\epsilon$. The primary plateaus at $\omega_p/\omega=1$ ($\epsilon\lesssim 0.28$) and $-2$ ($\epsilon\gtrsim 0.58$) correspond to trapping by $K_1$ and $K_4$, respectively. In contrast, the plateau $\omega_p/\omega=0$---no deterministic rotation---indicates neither wave can dominate over the other when $\epsilon$ is near $\epsilon^*\simeq 0.29$. The higher-order plateaus are significantly narrower and thus difficult to visualize experimentally due to finite modulation strength increments ($\delta\epsilon\simeq 0.01$). Nevertheless, numerical simulations (red line) can resolve the higher-order plateaus (see Fig.~\ref{fig:wave_transition}b and SI). These simulations show that the widths of these Shapiro steps depend on $\omega$, but the plateau values are fixed and can be reproduced via a Farey tree construction (see SI) \cite{1999-Reichhardt-Phase-PRLa}.

\subsection*{Directional locking in synthetic space}
Next, we analyze the ring's dynamics driven by multiple waves in a higher-dimensional synthetic space. Note that analogous techniques to engineer synthetic spaces have facilitated the study of higher-dimensional topological quantum matter and photonics \cite{2012-Boada-Quantum-PRL,2012-Edge-Metallic-PRL,2012-Kraus-Topological-PRLb,2013-Kraus-Fourdimensional-PRL,2013-Kraus-Fourdimensional-PRL,2013-Verbin-Observation-PRL,2015-Manai-Experimental-PRL,2016-Lohse-Thouless-NP,2016-Nakajima-Topological-NP,2018-Yuan-Synthetic-Oa,2019-Ozawa-Topological-NRP}. To this end, we construct the phase lag $n^{(i)}(t)=|K_i[\omega_i t-\theta(t)]|/(2\pi)$ ($i=1,4$) between the ring and each wave, scaled by the corresponding wavelength. Figure~\ref{fig:wave_transition}c plots $n^{(4)}$ vs $n^{(1)}$ from the same data in Fig.~\ref{fig:wave_transition}a. At $\epsilon=0.21$, the {\it phase} trajectory aligns with the $n^{(4)}$ axis. In contrast, at $\epsilon=0.68$, the trajectory aligns with the $n^{(1)}$ axis. For the intermediate $\epsilon=0.46$, the trajectory traces a zigzag path through the 2D lattice defined by $\{n^{(1)},n^{(4)}\}$, closely resembling particle trajectories driven across 2D real-space periodic potentials \cite{2002-Korda-Kinetically-PRL}.

As in real space, we define the angle $\Phi=\tan^{-1}[\Delta n^{(4)}/\Delta n^{(1)}]$ to quantify the direction of a trajectory in the synthetic space, where $\Delta n^{(i)}$ ($i=1,4$) is the total phase displacement. As $\epsilon$ increases, $\Phi$ exhibits distinct plateaus that form a Devil's staircase, closely mirroring the hallmark of kinetic locking in 2D real space (Fig.~\ref{fig:wave_transition}d). The primary plateaus at $\Phi=\pi/2$ and $0$ correspond to locked directions along $n^{(4)}$ and $n^{(1)}$ axes, respectively. In contrast, the plateau at $\Phi=\pi/4$ ($0.30\lesssim\epsilon\lesssim 0.43$) indicates a locked diagonal direction in synthetic space. Notably, the ring's mean rotation speed and the synthetic-space direction are related by $\tan\Phi=(\omega_p/\omega+2)/(2-2\omega_p/\omega)$ (see SI), which establishes an explicit mapping from real-space mode locking (Fig.~\ref{fig:wave_transition}b) to synthetic-space kinetic locking (Fig.~\ref{fig:wave_transition}d). Thus, the mode-locking Shapiro steps and kinetic-locking Devil's staircase exhibited by the colloidal ring are simply distinct manifestations of the same nonlinear dynamics driven by multiple time-dependent forces. 

Notably, our experimental conditions differ from typical real-space kinetic locking studies where the potential is fixed and the forcing direction varies \cite{2002-Korda-Kinetically-PRL}. Here, increasing $\epsilon$ raises potential barriers within $K_4$, thereby hindering hops along the $n^{(4)}$ axis, while barriers in $K_1$ remain unchanged. Further increases in $\epsilon$ cause the barriers along the $n^{(4)}$ axis to first become comparable to, and then surpass, those along the $n^{(1)}$ axis. Additionally, viscous drag drives hops diagonally in the $\{n^{(1)},n^{(4)}\}$ space because $K_i\omega_i=M\omega$ holds for both waves. Despite these differences, the rational plateaus of locked directions in the synthetic space and real space are both determined by the symmetry of a square lattice and thus are of identical values \cite{1999-Reichhardt-Phase-PRLa,2004-Gopinathan-Statistically-PRLa}. This new insight offers a geometric interpretation of both integer and fractional Shapiro steps in mode-locked dynamics \cite{2015-Juniper-Microscopic-NC,2025-Stikuts-Engineering-N}.

Comparing our observations with kinetically locked 2D clusters yields more insights. In Ref. \cite{2019-Cao-Orientational-NP}, overlapping sites between the cluster lattice and substrate lattice form superlattices for specific alignments; when the cluster is locked both orientationally and directionally, the same superlattice repeats. Conversely, when the cluster translates and rotates, superlattices of different symmetries form intermittently. In our system, single-wave driving beyond the critical speed repeatedly aligns the ring with a superlattice of potential traps in the wave (Fig.~\ref{fig:1}d). During inter-wave transitions, in contrast, the ring navigates between superlattices of traps belonging to different waves. Thus, the inter-wave transitions in our 1D cluster share similar characteristics with the inter-orientation transitions in locked 2D clusters.

\begin{figure}[h]
	\centering
	\includegraphics[width=.8\columnwidth]{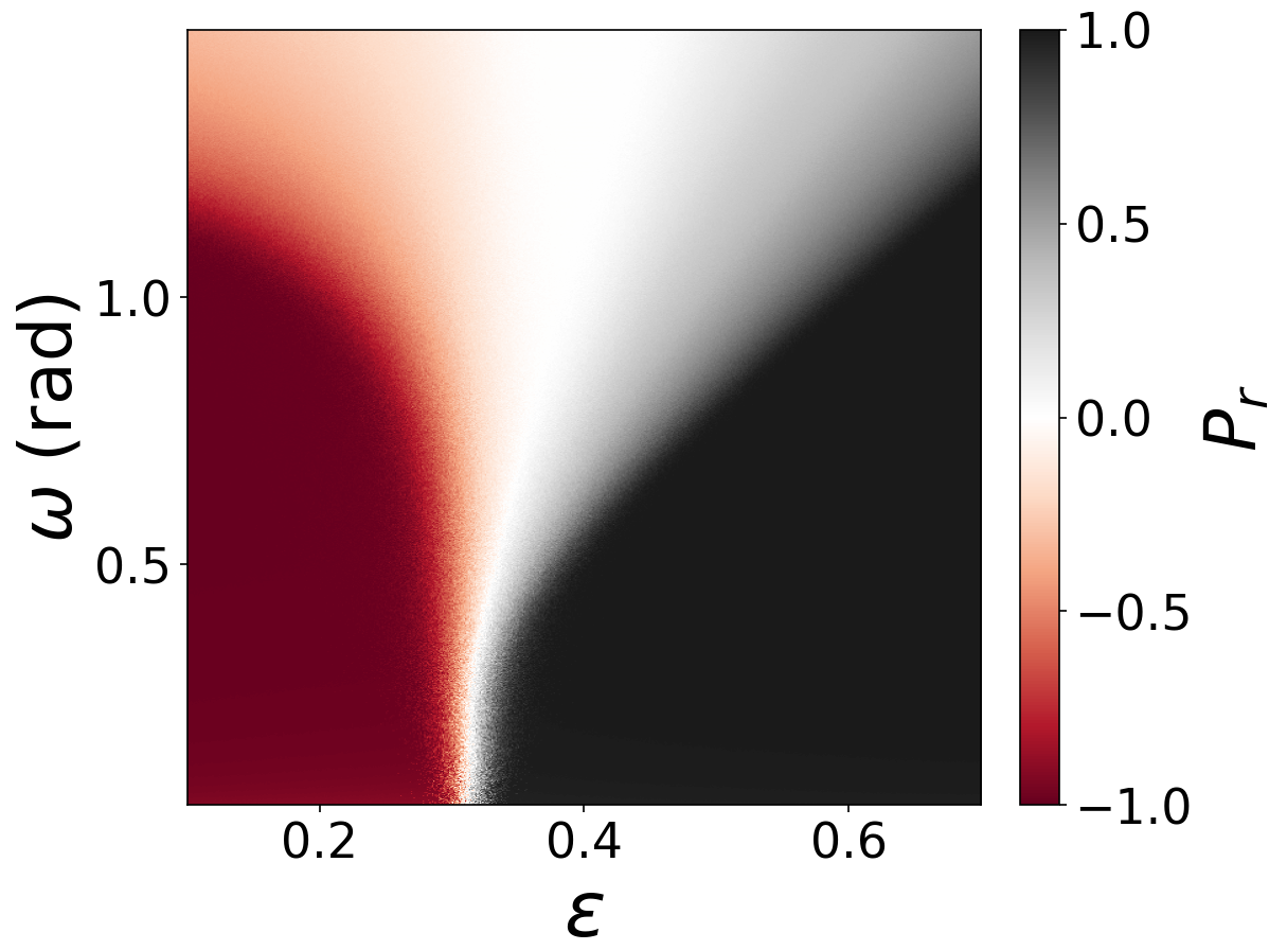}
	\caption{{\bf Competition between driving waves and viscous drag.} $P_r$ vs $\epsilon$ and $\omega$ from simulations. Red and black regions are dominated by $K_1$ and $K_4$, respectively.}
	\label{fig:dominance}
\end{figure}
Finally, we characterize the competition between driving waves and viscous drag by defining the ``dominance'' parameter $P_r=[\Delta n^{(1)}-\Delta n^{(4)}]/[\Delta n^{(1)}+\Delta n^{(4)}]$. Values of $P_r\to -1$ (where $\Delta n^{(1)}\ll \Delta n^{(4)}$) or $P_r\to 1$ (where $\Delta n^{(1)}\gg \Delta n^{(4)}$) indicate dominance of $K_1$ or $K_4$, respectively. When $P_r\to 0$ (where $\Delta n^{(1)}\simeq \Delta n^{(4)}$), the two waves are comparable; the driving forces of both waves may be superseded by viscous drag. Figure~\ref{fig:dominance} plots $P_r$ as a function of $\epsilon$ and $\omega$ from numerical simulations, revealing that $K_1$ dominates the bottom-left region while $K_4$ dominates the bottom-right. The central funnel-shaped region represents balanced competition between the driving waves at lower $\omega$ and dominance of viscous drag at higher $\omega$.

\section*{Discussion and conclusion}
We demonstrated that the driven dynamics of a 1D colloidal ring in a power-modulated optical tweezers array exhibit hallmarks of both real-space mode locking and synthetic-space (directional) kinetic locking. Specifically, when driven by multiple potential waves, the inter-wave transitions of the ring cluster give rise to both Shapiro steps in its mean rotation speed and a Devil’s staircase of its synthetic-space directions. This finding establishes an equivalence between real-space mode locking and synthetic-space kinetic locking, and identifies a unifying lattice symmetry-based mechanism that governs these two classes of nonlinear driven dynamics.

The superlattices formed between the cluster's constituent particles and the underlying potential traps reveal an energetic mechanism analogous to that governing kinetic locking in 2D clusters. This analogy confirms the feasibility of probing cluster locking phenomena in higher-dimensional energy landscapes using lower-dimensional experimental setups with time-dependent modulations. It is thus possible to engineer more than two driving potential waves to explore phenomena---such as transition loops and the kinetic locking of clusters in 3D synthetic space---that have been experimentally inaccessible to date.

From an application perspective, we showed that power modulation of laser tweezers affords unprecedented advantages for engineering potential waves to control the transport of particle clusters. The tunable wavelengths and wave speeds open numerous avenues for the experimental realization of diverse mode-locking models. For instance, counter-rotating waves can be employed as two periodic sliders to investigate solid lubricant dynamics \cite{2006-PhysRevLett.97.056101}. Additionally, the divisibility criterion can be exploited to selectively transport clusters of target particle sizes, enabling new cluster sorting techniques. These promising applications should be explored in future work.

\section*{Methods}
\subsection*{Experiment}
Polystyrene microspheres (nominal diameter \(D\simeq 2.5\;\mu\)m, Thermo Fisher Scientific) are suspended in an aqueous solution of $44$ mM hexaethylene glycol monododecyl ether (\(C_{12}E_6\)) and $2$ mM NaCl. The \(C_{12}E_6\) micelles provide temperature-tunable depletion-induced interparticle attractions \cite{2016-Ma-PhysRevE.94.042606,2019-Ma-10.1063/1.5091564,2021-Ma-10.1063/5.0059084,2023-Ma-10.1063/5.0146155}: at \(20^\circ\)C, the depletion effect is negligible, while at \(40^\circ\)C, the attraction strength (\(>7k_BT\)) suffices to stabilize rigid particle clusters. 

Approximately $50$ $\mu$L of the suspension is sandwiched between two parallel cover glasses at \(20^\circ\)C. The glass separation is slightly larger than \(D\) to confine the particles to a quasi-2D space. This sample cell is sealed with UV glue (Norland Optical Adhesive 65) and placed in a temperature-controlled incubator (UNO, Okolab, precision:\(0.1^\circ\)C) mounted on an inverted microscope (Ti2U, Nikon) stage. We employ a $100\times$ oil-immersion objective (N.A.=1.2) for bright-field imaging. Video frames are recorded at $20$ fps by a digital camera ($1920\times1280$ pixels; aca1920-155um, Basler). Particle positions and trajectories are extracted from the video frames with $20$ nm spatial resolution using a custom software.

A commercial optical tweezers system (Tweez 305, Aresis) generates multiple laser tweezers via time-sharing of a single $1064$ nm, $5$ W infrared laser beam. An acousto-optic deflector deflects the beam at $100$ kHz, which dwells at each pre-determined trap site for a few microseconds before moving to the next, enabling simultaneous trapping of multiple colloidal particles. We arrange \(M\) laser tweezers on a circular array (radius \(2.5\;\mu\)m). The power of each laser tweezers is set by the control software according to its current angular position. All tweezers positions are dynamically updated to rotate the tweezers array.

At \(20^\circ\)C, we use the laser tweezers to trap and assemble $7$ particles into a close-packed cluster (Fig.~\ref{fig:1}b). Then, we raise the sample temperature to \(40^\circ\)C, where strong depletion forces rigidly bound the particles. After the assembly, the colloidal cluster and the laser tweezers array are placed in a concyclic configuration (Fig.~\ref{fig:1}b).

\subsection*{Optical potential}
For an array of \(M\) power-modulated Gaussian laser beams, light intensity at the angular position \(\theta'\) contributed by the \(j\)-th beam (\(j=1,2,\dots,M\)) is \cite{2016-Egelhaaf-PhysRevA} (see SI for the detailed derivation):
\begin{equation}
I_j(\theta',t) = I_m\left[1+\epsilon\cos(L\theta_j)\right]\exp\left[-2R^2(\theta'-\theta_j)^2/w^2\right],
\label{eq03}
\end{equation}
where \(\theta_j(t)=\omega t+2\pi(j-1)/M\) is the angular position of the \(j\)-th beam center, \( I_m[1+\epsilon\cos(L\theta_j)] \) the light intensity at the beam center which scales linearly with laser power, and \(w\) the beam waist size.

The optical potential contributed by the \(j\)-th beam is:
\begin{equation}
u_j(\theta,t)=-\kappa I_m \frac{w \sqrt{2 \pi}}{4 \pi R} \left[1+\epsilon\cos(L\theta_j)\right]\sum_{k=-\infty}^{\infty} B_{k}C_{k} e^{i k(\theta-\theta_j)},
\label{eq04}
\end{equation}
where \(\kappa\) is the effective polarizability coupling light intensity to potential energy, and \( D(\theta,\theta') \) is a normalized weight function. Below, we present results using a Gaussian-form \(D(\theta,\theta')\). For improved accuracy, a semi-circle-form \(D(\theta,\theta')\) is used instead in numerical simulations (see SI). 

The total potential $u(\theta, t)$ experienced by a particle is the superposition of contributions from the \(M\) beams, each decomposed into a Fourier series (see SI):
\begin{equation}
\begin{aligned}
	u(\theta, t) &= -A_0 \sum_{m = -\infty}^{\infty} B_{mM}C_{mM} e^{i mM (\theta-\omega t)} \\
	&\quad - \frac{A_0 \epsilon}{2} \sum_{m = -\infty}^{\infty} \bigg[ \\
	&\qquad  B_{mM+L}C_{mM+L} e^{i [(mM+L) \theta - mM\omega t]} \\
	&\qquad  + B_{mM-L}C_{mM-L} e^{i [(mM-L) \theta - mM\omega t]} \bigg],
\end{aligned}
\label{eq:potential_fourier_final}
\end{equation}
where $m$ is the Fourier index and \(A_0=\kappa I_m w M \sqrt{2\pi}/(4\pi R)\) is a constant potential determined by laser beam parameters. The combined factor \( B_k C_k = \exp(-Ck^2) \) with $k=mM,mM\pm L$ and $C = w^2/(8 R^2) + \pi^2/(4 N^2 \ln 2)$ strongly suppresses higher-order ($|m|$) or larger-wavenumber ($k$) modes, leaving the zero- (\(m=0\)) and first-order (\(|m|=1\)) modes dominating the total potential as described by Eq.~(\ref{eq:potential}). The coefficients \(\{b_i\}\) are:
\begin{align}
b_1 &= 2A_0 \exp(-CM^2) \nonumber \\
b_2 &= A_0 \exp(-CL^2)  \nonumber\\
b_3 &= A_0 \exp\left[-C(M+L)^2\right] \nonumber \\
b_4 &= A_0 \exp\left[-C(M-L)^2\right].
\label{eq:b_coeffs}
\end{align}

\subsection*{Brownian dynamics}
The ring's dynamics are governed by the overdamped Langevin equation:
\begin{equation}
	\gamma R \dot{\theta}=-\frac{1}{R} \frac{\partial U(\theta, t)}{\partial \theta}+\eta(t),
	\label{eq:langevin}
\end{equation}
where $\gamma$ is the friction coefficient and $\eta(t)$ denotes the Brownian noise. The noise satisfies $\langle\eta(t)\rangle=0$ and $\langle\eta(0)\eta(t)\rangle = 2\gamma k_BT\delta(t)$, where $\delta$ is the Dirac delta function, $k_B$ Boltzmann's constant, and $T$ the temperature. When the noise is negligible, Eq.~(\ref{eq:langevin}) simplifies to a deterministic equation:
\begin{align}
	\gamma R^2\dot{\theta} =  &\sigma(K_1)a_1\sin[K_1(\omega_1 t-\theta)]\nonumber \\
	- &\sigma(K_2)a_2\sin(K_2\theta)\nonumber \\
	+&\sigma(K_3)a_3\sin[K_3(\omega_3 t-\theta)]\nonumber \\
	+&\sigma(K_4)a_4\sin[K_4(\omega_4 t-\theta)],
	\label{eq:eom}
\end{align}
where $\{a_i\}$ are wavenumber-dependent amplitudes (Table~\ref{tab:amplitudes}).  

When only one wave $K_i$ exists, the ring's steady-state angular speed $\omega_p$ is determined using the trial solution $\theta(t)=\omega_p t - \psi$, yielding $\omega_p=\omega_i$, indicating the ring is trapped by $K_i$. The phase $\psi=\sin^{-1}[\gamma R^2M\omega/(a_iK_i)]/K_i$ requires $|\gamma R^2M\omega/(a_iK_i)|\leq 1$; this defines the critical tweezers speed $|\omega^*|=a_iK_i/(\gamma R^2 M)$, beyond which the trial solution becomes invalid.

\section*{Data Availability}
The authors declare that all data supporting the findings of this work are available within the paper and its Supplementary Information files or available from the authors upon reasonable request. Source data are provided in this paper.

\section*{Acknowledgments}
We thank Xin Cao, Penger Tong, and Qi-Huo Wei for helpful discussions. MH and XM thanks the National Natural Science Foundation of China (Grant No. 12274195), National Key Research and Development Program of China (Grant No. 2022YFA1405002), Department of Science and Technology of Guangdong Province (Grant No. 2021QN02C382). PYL thanks the National Science and Technology Council of Taiwan (Grant No. 113-2112-M008-018-MY2).

\bibliography{ref}

\end{document}